# Correlating Nanoscale Structure with Electrochemical Property of Solid Electrolyte Interphases in Solid-State Battery Electrodes

*Jimin Oh, Gun Park, Hongjun Kim, Sujung Kim, Dong Ok Shin, Kwang Man Kim, Hye Ryung Byon, Young-Gi Lee, and Seungbum Hong\**

*J. Oh, G. Park, Dr. H. Kim, Prof. S. Hong*
Department of Materials Science and Engineering, Korea Advanced Institute of Science and Technology (KAIST), Daejeon 3414, Republic of Korea.
E-mail: seungbum@kaist.ac.kr

*Dr. S. Kim, Prof. H. R. Byon*
Department of Chemistry, Korea Advanced Institute of Science and Technology (KAIST), Daejeon 34141, Republic of Korea.

*Prof. S. Hong*
KAIST Institute for the NanoCentury, KAIST, Daejeon 34141, Republic of Korea.

*J. Oh, Dr. D. O. Shin, Dr. K. M. Kim, Dr. Y.-G. Lee*
ICT Creative Research Laboratory, Electronics and Telecommunications Research Institute, Daejeon 34129, Republic of Korea.

Here, we correlate the nanoscale morphology and chemical composition of solid electrolyte interphases (SEI) with the electrochemical property of graphite-based composite electrodes. Using electrochemical strain microscopy (ESM) and X-ray photoelectron spectroscopy (XPS), changes of chemical composition and morphology (Li and F distribution) in SEI layers on the electrodes as a function of solid electrolyte contents are analyzed. As a result, we find a strong correlation between morphological variations on the electrode, Li and F distribution in SEI layer, and Coulomb efficiency. This correlation determines the optimum composition of the composite electrode surface that can maximize the physical and chemical uniformity of the solid electrolyte on the electrode, which is a key parameter to increase electrochemical performance in solid-state batteries.

# 1. Introduction

The technology of lithium-ion batteries has developed over time and now features in applications from tens of watt-hour (Wh) small home appliances and smartphones to tens of kWh and more electric vehicles (EVs) and energy storage systems (ESSs)[1]. Such large-scale lithium-ion battery applications demand superior safety characteristics. To improve the safety of lithium-ion batteries, solid-state lithium-ion batteries employing solid electrolytes have been researched for the last several decades; the advantages of solid electrolytes include non-flammability and a wide electrochemical window of stability.

It should be noted that cell safety can be improved through the use of a non-liquid electrolyte, thus removing the risk of electrolyte leakage and ignition. In addition, the advantageous mechanical properties of solid electrolytes suppress the formation of lithium dendrites, which safely provide high energy density cells featuring a lithium-metal electrode[2]. Different types of solid-state batteries have been reported, including several candidates that employ oxide electrolytes[3], sulfide electrolytes[4], solid polymer electrolytes[5], and solid-liquid composite electrolytes[6]. Despite these merits, owing to the low ionic conductivity of solid electrolytes and a high interfacial resistance between active materials and solid electrolytes, the application of solid-state batteries has been limited.

To aid researchers in overcoming the disadvantages of solid-state batteries, analysis techniques are required that enable the quantification of the ion mobility and diffusion coefficient as well as the visualization of local ion concentrations and distributions in solid electrolytes[7-9].

At the nanometer and atomic level, atomic force microscopy (AFM) is a significant analytical scanning probe microscopy (SPM) technique. AFM is a high-resolution imaging technique that analyzes the surface morphologies of samples. It can extract information on the generation of surface interlayers as well as the mechanical and electrical properties of the electrodes[9-11]. Electrochemical strain microscopy (ESM), one of functional AFM modes, has been recently developed to visualize surface lattice variation induced by electrical potentials between a tip and a sample. The ESM, electrically-biased contact AFM mode, and its derivative techniques are utilized for the analysis of local ion concentrations and lattice variations on the unit cell down to the nanometer level[12, 13]. Alikin *et al.* probed the local diffusion coefficients and ion concentrations of particles of lithium-ion battery cathode materials, $LiMn_2O_4$, with different sizes[14]. Their quantitative analysis of the ionic mobility and concentration unveiled the dynamic behavior of ion migration and relaxation and the change in ion concentration profiles.

For the characterization of solid-state batteries, it is necessary to note that the analysis of surface chemical distributions and concentrations after electrochemical reactions provides powerful insights into the aging degradation mechanisms and capacity properties of cells[15]. Using the ESM method, Wang *et al.* revealed the movement of ions between the grain and grain boundaries at different sites of a sodium (Na) super ionic conductor (NASICON)-structured $Li_{1.5}Al_{0.5}Ge_{1.5}(PO_4)_3$ (LAGP) solid electrolyte[16]. Furthermore, Strelcov *et al.* have developed functional electrochemical probing methods applicable to liquid–solid interfaces, and, for the current trend in the development of all-solid-state batteries, have determined optimal SPM techniques for a multitude of complex cross-coupled physical and chemical phenomena in liquid-free environments[17].

In this paper, we present the electrochemical effects of solid electrolytes (e.g., lithium silicon titanium phosphate (LSTP)) on composite electrodes via visualization of topographic variations and analysis of the chemical composition on a nanometer scale of the solid electrolyte interphase (SEI) layers. The chemical distributions and morphology characterizations of the SEI layers provide insights for researchers to develop high energy density and safety-enhanced lithium-ion batteries, since correlations between Coulomb efficiency and chemical composition help us to understand the mechanisms governing ion mobility in solid-state batteries. The mapping of the lithium-ion distributions on the surface of the SEI layers is of specific interest in practical surface analyses of electrode degradation in batteries[8, 13, 18]. In addition, information concerning the chemical concentration and distribution in SEI layers on the electrodes is necessary to determine the presence and extent of unnecessary and undesirable electrochemical reactions

## 2. Results and discussion

## 2.1 Visualization of Li Distribution as a function of Lithiation on the Graphite-based Electrodes

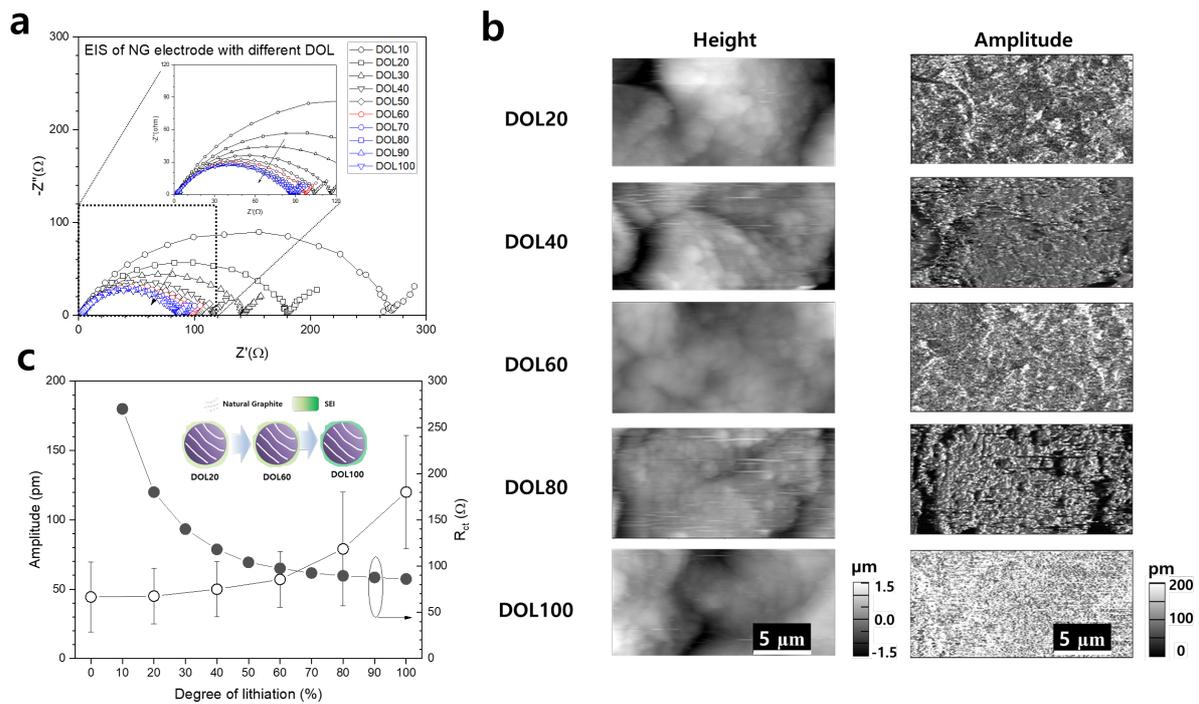

**Figure 1.** (a) EIS results of NG electrode with different DOLs, (b) topographic/ESM images and (c) plot of amplitude, and charge transfer resistance at different DOL using conventional NG electrodes.

In order to visualize Li distribution on composite electrode, we conducted ESM imaging of conventional natural graphite (NG) electrode as a reference. **Figure 1** shows the electrochemical impedance spectroscopy (EIS) results, topography and ESM images and relationship between ESM and charge transfer resistance of conventional NG electrode.

The electrodes consist of NG/SBR-CMC/Super-P with five different degrees of lithiation (DOL) states where DOL20, DOL40, DOL60, DOL80 and DOL100 represent 20 %, 40 %, 60 %, 80 % and 100 % of lithiations, respectively (Figure S1). Five different states of samples were prepared and lithiated at 0.1 C-rate (37.2 mAh $g^{-1}_{NG}$). Each charge transfer resistance was observed by EIS.

The surface morphology shows granular-shape NG particle with diameter of 10 ± 1.35 μm. In the height images, most NG surfaces are covered with various size of turtle-shell-shaped agglomerates that are derived from electrolyte decompositions under 0.1 V vs Li/Li$^+$, depending on DOL.

ESM images show the spatial distribution of Li ion on the surface of Li ion battery materials[10, 19, 20]. Therefore, the average ESM amplitude is expected to linearly scale with DOL[21]. As DOL increases, Li distribution (bright region at amplitude map) was observed from boundary of NG particles (DOL20) to all area on the electrode (DOL100). However, the amplitude increased in a nonlinear fashion in Figure 1c. In order to understand this discrepancy, we analyzed the charge transfer resistance as a function of DOL.

From the EIS results, we observed that the charge transfer resistance decreases as the lithiation increases in a nonlinear fashion. Similarly, the ESM amplitude increases as DOL increases in a nonlinear fashion. In the low DOL regime, the amount of Li intercalation will be hindered by high charge transfer resistance whereas in the high DOL regime, the Li intercalation will exponentially increase due to low charge transfer resistance.

ESM amplitude image of DOL20 showed bright region near NG particles due to the high charge transfer resistance while that of DOL100 showed uniform bright contrast over the whole region on the electrode due to the low resistance. Therefore, one can correlate the Li ion distribution on the electrodes at nanoscale with the electrochemical reactions at much larger scale.

**2.2 Visualization of Solid Electrolyte Effects on the Composite Electrodes**

Solid electrolyte affects the lithium-ion diffusivity and storage capacity in the composite electrodes, which, in turn, determine the entire electrochemical properties at solid-state batteries. Most of SEI is known to be formed at the first cycle under 1 V through mixing of decomposed lithium salt and solvent. However, the formation mechanism of SEI in the vicinity of composite electrode mixed with solid electrolyte has remained largely unknown.

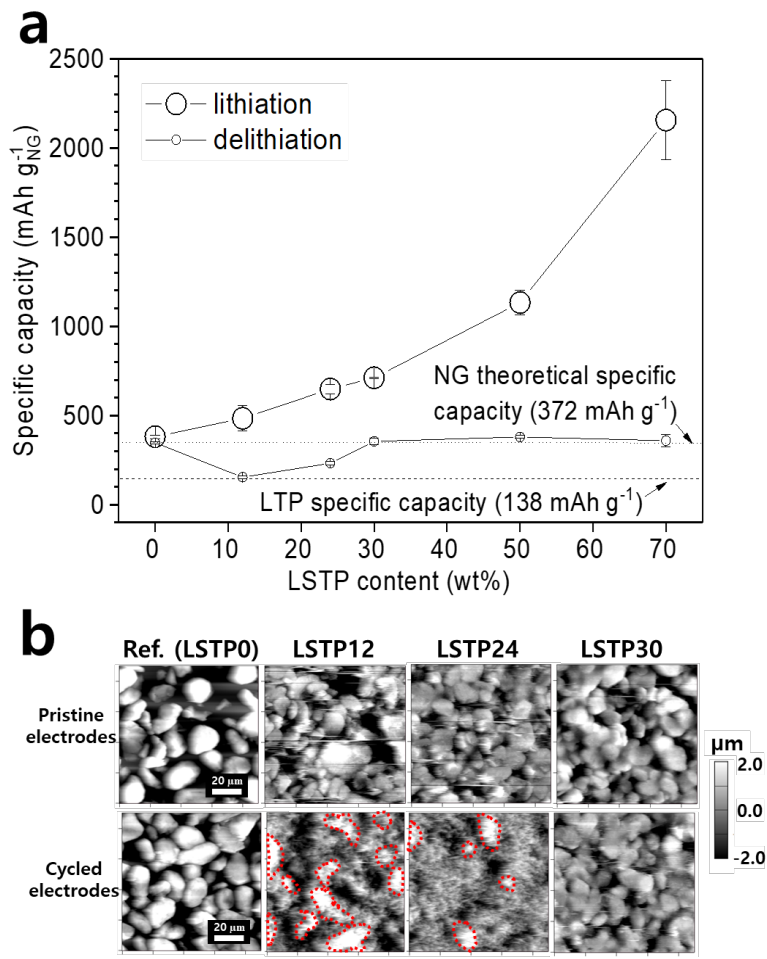

**Figure 2.** (a) Plot of specific capacity of composite electrodes after lithiation and delithiation as a function of LSTP content. (b) AFM topography images of pristine and cycled electrodes as a function of LSTP content. The red dashed lines represent the boundary of regions higher than a certain threshold (2 μm) for LSTP12 and LSTP24.

To investigate the effect of solid electrolyte on the composite electrode, we fabricated composite electrodes, composed of NG/LSTP/SBR-CMC/Super-P with different NG/LSTP ratios. **Figure 2** shows lithiation-delithiation specific capacities and height images of pristine and cycled electrodes as a function of LSTP content. **Table 1** summarizes the values of lithiation-delithiation specific capacities, irreversible specific capacities, and Coulomb efficiency for the composite electrode samples.

**Table 1.** Electrochemical properties of composite electrode after cycling at 0.1 C in the first cycle. 5 different cells were prepared for each sample case.

| Sample | Average lithiation specific capacity ± deviation [mAh·g$^{-1}$] | Average delithiation specific capacity ± deviation [mAh·g$^{-1}$] | Average irreversible specific capacity ± deviation [mAh·g$^{-1}$] | Coulomb efficiency ± deviation [%] |
|---|---|---|---|---|
| Ref. (LSTP0) | 355±36 | 326±29 | 29±7 | 91.7±1.74 |
| LSTP12 | 406±149 | 110±53 | 296±96 | 26.5±8.64 |
| LSTP24 | 633±42 | 184±57 | 448±33 | 29.1±9.67 |
| LSTP30 | 660±58 | 331±51 | 330±49 | 49.9±3.16 |
| LSTP50 | 952±251 | 257±131 | 695±120 | 26.1±8.94 |
| LSTP70 | 2155±223 | 359±35 | 1796±188 | 16.7±0.10 |

As the LSTP content increases, the lithiation specific capacity showed an overall increasing trend when lithiated whereas it fluctuated down and up and saturated at the level of theoretical specific capacity of NG for more than 30 wt % of LSTP content when delithiated (as shown in Figure 2a and Figure S2).

The reason for overall increase in the lithiation specific capacity is the significant increase of the specific surface area. The specific surface area of the electrode increases with addition of LSTP from Equation (1),

$$\beta_{\text{elec}} = M_g \beta_g + M_{\text{LSTP}} \beta_{\text{LSTP}} \tag{1}$$

where $M_g$, $M_{\text{LSTP}}$ denote mass fractions of NG and LSTP, respectively. $\beta_g$, $\beta_{\text{LSTP}}$ are specific surface area of NG (10 m$^2$ g$^{-1}$) and LSTP (60-70 m$^2$ g$^{-1}$), respectively. The specific surface area with higher LSTP amount in the electrode induces more electrochemical reactions, resulting in higher electron consumptions.

Figure 2b shows topographical maps in pristine and cycled electrodes as a function of LSTP contents up to 30 wt%. NG particles are clearly seen in all of the pristine electrodes. After initial cycle, NG particles remained well distinguished in the LSTP0 and LSTP30 electrodes, while they became obscure in LSTP12 and LSTP24. The lower Coulomb efficiency, the ratio of delithiation and lithiation specific capacity, induces a significant change of surface topography. Also, delithiation capacity characteristics affect the deposition surface pattern near the active material on the electrodes.

In order to understand the change of topography in Figure 2b for cycled composite electrodes of LSTP12 and LSTP24, we analyzed the contour of pixels higher than a certain threshold in the height image and colored it red. Clusters of white regions surrounded by red contour could be identified, which we presume to be non-uniformly deposited lithium compounds. The compounds consume excessive lithium ions and electrons, which leads to an increase of irreversible specific capacity and decrease of Coulombic efficiency.

The reason we did not analyze the topography images of composite electrodes with LSTP contents more than 30 wt% is because NG particles were mostly covered with LSTP, making our comparison between pristine and cycled electrodes difficult (Figure S3).

## 2.3 Quantitative Analysis of Li Distribution and Concentration in SEI layers

ESM imaging was conducted to understand the microscopic origin of topographic change induced by a sequence of lithiation and delithiation as a function of LSTP content. When conducting ESM, one needs to decouple the electromechanical response from the electrostatic effects induced between the cantilever and samples by voltage-modulated force spectroscopy [22-25]. Firstly, we apply bias voltage to the tip, which makes the electric field penetrate near the sample surface down to a depth of 50 to 100 nm, where most of SEI layer is concentrated. Additionally, a perfect SEI is pure ionic conductor, which leads to electronic charge injection and trapping in the SEI layer. This can induce a large electrostatic effect in the acquired ESM signal.

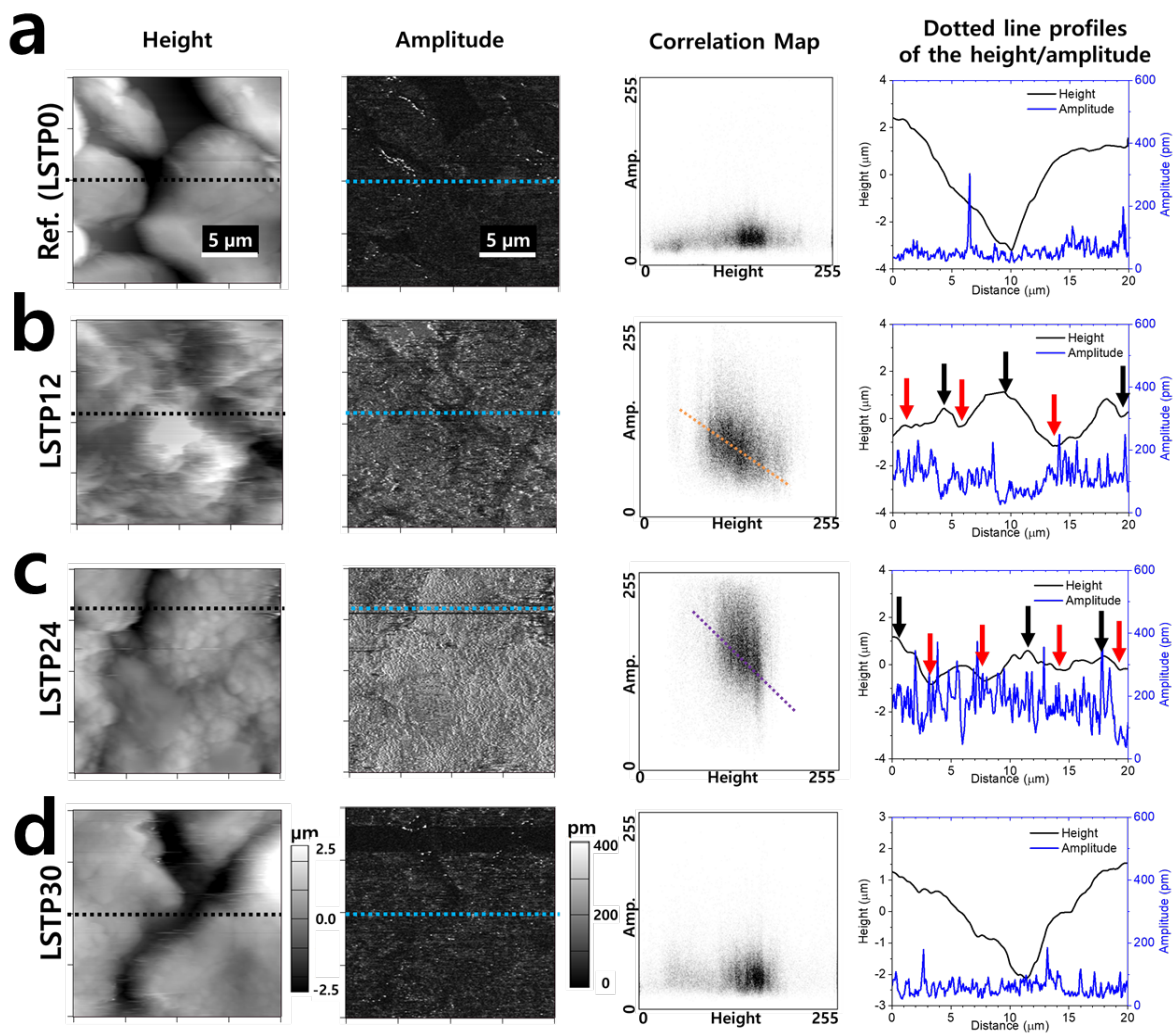

**Figure 3**. Heights, amplitudes, correlation maps and line profiles of a) Ref. (LSTP0), b) LSTP12, c) LSTP24, and d) LSTP30 electrodes.

However, a thick and imperfect SEI layer can be considered as a mixed conductor because it displays a low Coulomb efficiency. Therefore, the ESM signal from a mixed conductor can be attributed mostly to the Vegard strain because the injected charges will have a very short trapping time on the surface. As such, we expect that our ESM signal is coming from the Vegard strain which is proportional to the Li ion concentration. In addition, we are aware of the fact that for more accurate analysis we need to measure the effective Vegard coefficient β. However, even without the knowledge of β, our analysis still holds true because we made relative comparison as a function of solid electrolyte content.

In order to prove our claim, we conducted Pearson correlation mapping between topography and ESM signals at the same position. In case of capacitive artifact in ESM images, there is often found strong linear relationship between topography and ESM signal. As such Pearson correlation plot is useful in checking the presence of capacitive artifact.

Other approach is to measure the second or higher contact resonance Eigen mode with the reduction of signal contributions from non-local electrostatic interactions between the sample and cantilever[24]. An alternative way has been suggested by Han *et al.*[25], where they removed the electrostatic effects by increasing the contact force that in turn increase the contact stiffness with minimal effect on the dynamic stiffness for the first contact resonance, which was proven by the cantilever dynamics modelling. As such, we choose the high contact force (~250 nN) to reduce the electrostatic effects between the sample and the cantilever.

In **Figure 3**, height and ESM amplitude images were compared side by side and their correlation was analyzed by visualization of Pearson correlation map. Images were obtained at 10 different positions of each sample and correlation map was obtained at 5 different positions from selected height, amplitude images to show the reproducibility of our methods (Figure S4). Pearson correlation plots in Figure 3 didn't show a perfect linear correlation but a complicated pattern between height and ESM signal, indicating the suppression of the capacitive artifact in the ESM images.

Line profiles of both height and ESM amplitude are shown to intuitively understand the correlation captured in correlation map analysis. The x and y axes represent height and ESM amplitude where the values were varied between 0 and 255 because we used 8-bit images to represent them in separate topography and ESM images. As such, 0 in height is -2.5 μm and 255 means 2.5 μm in x axis, and 0 in ESM amplitude means 0 pm, and 255 means 400 pm.

In the case of reference NG electrode, NG particles were clearly revealed in the topography image. In the meantime, ESM amplitude image showed an overall dark contrast with a faint resemblance of topographic features. Furthermore, bright regions were observed near some of the NG particle interfaces.

Based on these two images, Pearson correlation map is obtained to elucidate the relationship between height and ESM amplitude. The result shows statistically a narrow deviation of the amplitude with low value in a wide range of height. In the meantime, the height distribution was concentrated at a height slightly larger than the middle point.

The narrowly distributed and low ESM amplitude, regardless of the widely spread-out height, implies that thin active SEI layer uniformly covers the NG particles over the region of interest. Moreover, the concentrated distribution of height indicates that the surface of NG particles is relatively flat, as they occupy the majority of the surface where the flat NG particles formed by cold press during the fabrication are presented.

For further understanding the correlation, we extracted a profile along an arbitrary line over 20 μm distance in both height and ESM amplitude images (see Figure 3a). As a result, we could see a big valley in the height profile with an overall V-shape, representing the junction between two NG particles, which means the presence of a gap between NG particles. At the position near 10 μm, a boundary between the two NG particles was clearly observed.

However, the ESM amplitude fluctuated between 20 and 50 pm regardless of the position along the line profile. The line profile analysis supports our finding via Pearson correlation map analysis where the ESM amplitude shows statistically narrow deviation over a wide range of height.

In case of LSTP12, NG particles were not clearly identified in the topography images. The adsorbates appeared to be unevenly covering the surface area of NG particles. In the meantime, ESM amplitude showed overall gray contrast with some noise and dark contrast along some valleys coinciding with the part of the boundaries in the topography. However, there was no strong correlation between the ESM amplitude and topography images.

Our finding suggests that with addition of LSTP the lithiation occurs more uniformly on the surface regardless of the local topographic variation. Correlation map shows a wide deviation of the amplitude in a narrower range of height than Figure 3a. In the meantime, height data was distributed around the middle point. One possible way to understand this pattern in correlation map is to presume that SEI layer fully covered the valleys between NG particles, which results in a relatively smooth and flat surface whereas the active volume of Li-ions varies more drastically. Furthermore, weakly negative correlation between height and ESM amplitude was observed, which may explain why we observed dark contrast along some in the topographic image.

In order to elucidate the correlation further, we also extracted a profile along an arbitrary line over 20 μm distance in both height and ESM amplitude images (see Figure 3b). As a result, the amplitude shows high value where the height is relatively low (see red arrows), while the amplitude shows low value where the height is relatively high (see black arrows). This negative relationship means that SEI layer preferentially deposited near lower height regions, as shown in line profiles. The line profile analysis supports our correlation results where the ESM amplitude shows wide deviation over a relatively narrow distribution of height.

In case of LSTP24, some NG particles are clearly observed. The turtle-shell-shaped adsorbates are also clearly revealed on the NG surfaces. In the meantime, ESM amplitude image showed an overall bright contrast without resemblance of the topographic image, indicating that SEI layer covers uniformly over the electrode. Image correlation map shows a wider distribution of the ESM amplitude in a narrower range of height when compared with LSTP12 in Figure 3b.

Moreover, the weakly negative relationship between height and ESM amplitude is also observed. However, the slope of the negative relationship is steeper in LSTP24 (purple color dotted line) than in LSTP12 (orange color dotted line). This can be explained by the fact that the electrode is more fully covered in the case of LSTP24 when compared with LSTP12, leading to less variation in the surface height but more variation of active Li volume, which in turn results in more variation in ESM amplitude in the valleys between NG particles for the fully covered case.

The widely spread-out ESM amplitude at the middle point regardless of the narrowly distributed height implies that thick SEI layer covers both NG particles and a valley of their particles. In order to understand the correlation further, we conducted a profile analysis along an arbitrary line over 20 μm distance in both height and ESM amplitude images (see Figure 3c).

As a result, we could see a small valley in the height profile. However, the ESM amplitude fluctuated between 40 and 90 pm regardless of the position along the line profile. Similar with the result of LSTP12, the amplitude shows high value where the height is low (see red arrows), while the amplitude reveals low value where the height is high (see black arrows). The line profile analysis also supports our correlation results.

In the case of LSTP30 electrode, NG particles were clearly observed in the topography image. In the meantime, ESM amplitude image contains overall dark contrast with a weak resemblance of topographic features. Furthermore, brightest regions were observed on the surface of NG particles. Based on these two images, image correlation maps are obtained to confirm the relationship between height and ESM amplitude. The result shows a narrow deviation of the amplitude in a wide range of height. In the meantime, the height distribution was concentrated at a height slightly larger than middle region.

The narrowly distributed ESM amplitude regardless of the spread-out height implies that thin SEI layer homogeneously covers the NG particles over the region of interest. Furthermore, the concentrated distribution of height indicates that the surface of NG particles is relatively flat, as the surface is occupied with SEI compounds. For further understanding the correlation, we extracted a profile along a line over the same distance in both height and ESM amplitude (see Figure 3d).

As a result, we could see a big valley in the height profile with an overall V-shape, similar to the result shown in Figure 3a. The ESM amplitude fluctuated between 10 and 40 pm regardless of the position along the line profile. The line profile analysis confirms our finding via correlation analysis where the ESM amplitude shows narrow deviation over a wide range of height.

In addition to measurements of SEI properties on the cycled electrodes, we measured EIS of pristine and single-cycled states (Figure S5) and force-distance curves on the surface of cycled electrode (Figure S6) in reference electrode (LSPT0) and LSTP30. In pristine state of both electrodes, charge transfer resistances, calculated from the measured semi-circle in EIS results, are similar (85 Ω in Ref. vs. 70 Ω in LSTP30). After 1 cycle, charge transfer resistance in reference electrode is reduced to 65 Ω, due to generation of thin highly ionic conductive SEI layer, while charge transfer resistance in LSTP30 increased to over 150 Ω, due to the increased SEI thickness, which led to a low Coulombic efficiency.

This result supports increasing irreversible capacity in LSTP30 electrode than one in reference electrodes. In this regard, effect of SEI thickness from tip-sample distance obtained by AFM indentation is significant in Figure S6, Supporting Information. The total thickness of 2 layer-SEI is measured to be 45 nm for the reference electrode, while the total thickness of 3 layer-SEI is measured to be 298 nm for the LSTP30 electrode. It is noted that 45 nm of reference electrode (graphite) is within the range of SEI thickness between 20 nm and 80 nm measured by other research group[26]. This result also supports increasing irreversible capacity in LSTP30 electrodes.

**2.4 In-depth Chemical Composition Analysis**

Through height and ESM amplitude analysis in nanoscale, we found the correlation between the overall morphology of SEI layer and the distribution of Li ions in SEI layer as a function of LSTP content. Here, in order to have deeper understanding why the morphology of SEI severely changes in cases of LSTP12 and LSTP24, we utilized depth resolved XPS experiments using Ar ion etching to identify chemical compounds in SEI layers. In order to obtain surface chemical compositions, scanning electron microscopy (SEM) and energy-dispersive X-ray spectroscopy (EDS) results of cycled electrodes on the surface are included (Figure S8).

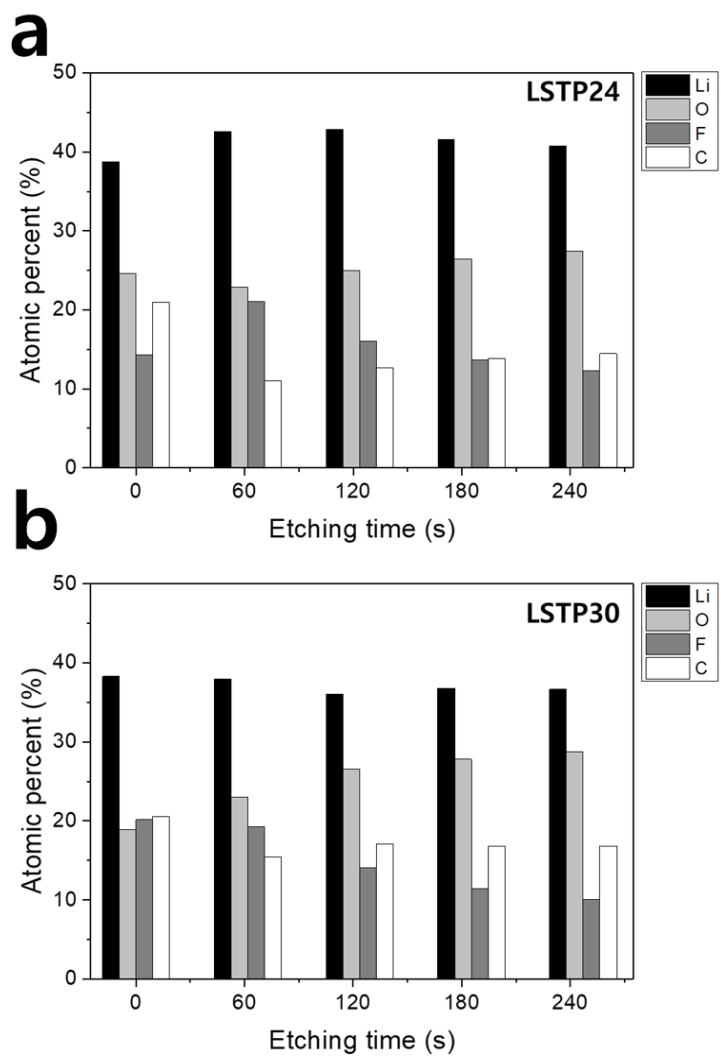

**Figure 4**. Atomic fractions of four elements of cycled electrodes in (a) LSTP24 and (b) LSTP30 in depth direction from XPS measurements.

Among four different LSTP composite electrodes, we chose LSTP24 and LSTP30 that showed distinct difference in both height and ESM amplitude distributions. Here, we assumed that different chemical compounds may form in SEI layer as we increase the LSTP amount in the composite electrode, which may account for drastic change in both morphology and Li ion distribution. In such a case, we expect to see peaks with different binding energies.

However, as shown in Figure S7, we observed that the peak positions in both LSTP24 and LSTP30 were identical in all binding energy ranges as a function of depth. Based on these results, we concluded that these changes occur even with the same chemical compounds in SEI layer.

Therefore, we checked whether the ratio of each chemical compound could determine the morphology and ion distribution change in SEI layer by analyzing its elemental ratio using XPS. Specifically, we analyzed the ratio of all four chemical elements (Li, O, F and C) in SEI layer as a function of etching time (see **Figure 4**).

The initial portion (38 %) of Li in LSTP24 is the same as LSTP30. However, as the depth increases, Li portion tends to increase and then decrease after passing through the maximum (43 %) in LSTP24. On the other hand, in LSTP30, as the depth increases, Li portion tends to decrease and then saturated after passing through the minimum (36 %).

Since the ESM signal mostly reflects the ionic information near the surface region, the higher ESM amplitude for LSTP24 in comparison to LSTP30 in Figure 3c, d can be explained by the fact that the integration of Li ions near the surface of LSTP24 is higher than that of LSTP30.

In case of F, its ratio directly represents the amount of LiF compounds in SEI layer, hence an excellent marker for its distribution. The fluorine element gradually decreased in LSTP30, while it fluctuated up to 20 % and down to 10 % in LSTP24. Based on our observation, LSTP24 shows higher intensity deviation compared to LSTP30, leading to more non-uniform passivation layer consisting mainly of LiF.

In addition, the distribution of oxygen and carbon elements can be linked to various compounds such as C=O, $CO_3^{2-}$, $Li_2O$, $TiO_2$, $Li_2CO_3$, and carbonaceous species, respectively as shown in Figure S7.

Table S1 summarizes the distribution of Li and F elements along the depth direction. The range of concentration was calculated by subtracting the minimum from the maximum and then normalized by the surface concentration. LSTP24 showed larger normalized range (0.106) than that of LSTP30 (0.059) even though the surface concentration was quite similar.

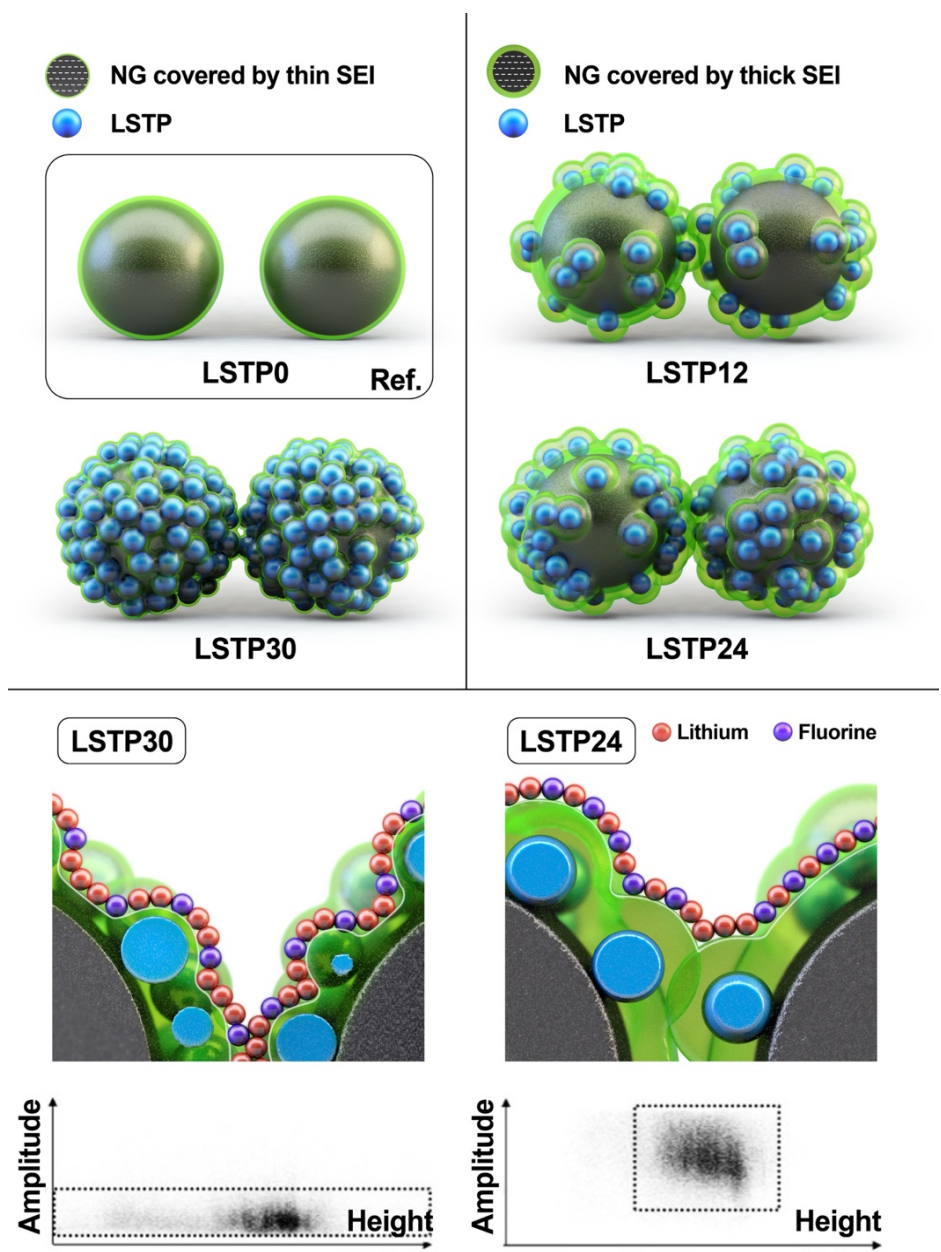

**Figure 5**. Schematic diagrams of SEI layers formed on the electrode depending on the solid electrolyte contents. Homogeneous depositions show a wide height and narrow amplitude whereas inhomogeneous depositions a narrow height and a wide amplitude.

The larger range can lead to more non-uniform Li distribution along the depth direction in SEI layer. Similarly, LSTP24 showed higher normalized range of fluorine (0.615) compared to LSTP30 (0.498), leading to more non-uniform distribution of fluorine element similar to Li.

By comparing the above results with irreversible specific capacity and Coulomb efficiency in Table 1, we found that nanoscale fluctuation of Li and F components in SEI layer is linked with macroscopic irreversible capacity and Coulomb efficiency.

In order to understand our results, we came up with a hypothetical model as depicted in **Figure 5**. The morphology and chemical homogeneity of SEI layer is affected by the component in direct contact with the liquid electrolyte. As such, for LSTP0 and LSTP30, the component in direct contact is either NG or LSTP, respectively.

The SEI properties (morphology and homogeneity) in conventional NG electrode are confirmed with different degree of lithiation. In particular, when the LSTP is almost uniformly covering the NG in LSTP30, the thickness of the SEI on the NG increases by the LSTP, but the increased SEI can also have homogeneity due to the uniform distribution of the LSTP.

Due to the electrochemical heterogeneity of NG and LSTP, it was confirmed that a stable SEI was formed in the electrode dominated by either NG or LSTP. On the other hand, for LSTP12 and LSTP24, the components in direct contact are both NG and LSTP. As a result, lithium-ion flux is affected by the distribution of LSTP. Interface between NG and LSTP accelerates the electrochemical reaction, which is exposed to the electrolyte in cases of LSTP12 and LSTP24.

## 3. Conclusion

We analyzed solid electrolyte interphases (SEI) layers derived from electrochemical effects of solid electrolyte on the graphite-based composite electrodes. We visualized Li ion distribution in SEI layers by DART mode ESM and characterized both Li and F ions using XPS. We found a strong correlation between morphological variations on the electrode, Li and F distribution in SEI layer, and Coulomb efficiency. Among the ratios of NG and electrolyte, 30 wt% LSTP sample showed the highest Coulomb efficiency with uniform surface morphology and chemical distribution in SEI layer on the composite electrode. Our findings provide insights into the physical and chemical properties of composite electrodes that can maximize the uniformity of SEI layer and hence minimize irreversible capacity, which is critical to increase charge-discharge performance of solid-state batteries.

## 4. Experimental Section/Methods

*Fabrication of Composite Electrodes*: All of the preparations were conducted under a dry room (dew point < −50 °C) and prepared via a tape casting method using a doctor blade with electrode slurry. The natural graphite (NG, BTR) (16 μm) and lithium silicon titanium phosphate (Jeong Kwan display) ($D_{50}$ = 250 nm, ion conductivity = 8 × $10^{-4}$ S $cm^{-1}$) were employed for an active material and a lithium-ion conductor, respectively. Carboxymethyl cellulose (CMC, Dai-Ichi Kogyo Seiyaku) and Styrene-Butadiene Rubber (SBR, ZEON) were prepared as a binder with a weight ratio of 1:4. The CMC-SBR binder was ionized in distilled water which its content was fixed at 2 wt % of the composite electrodes. Carbon black (Super-P, Timcal), an electron conductive material, was fixed at 0.5 wt%.

The weight ratio of NG and LSTP in composite electrodes was 97.5:0 (LSTP0), 85.5:12 (LSTP12), 73.5:24 (LSTP24), 67.5:30 (LSTP30), 47.5:50 (LSTP50) and 27.5:70 (LSTP70), respectively. The electrode slurry prepared under the above conditions was tape-cast on an 11 μm Cu film and dried at 100 °C for 12 h to remove the solvent and finally to obtain the electrode. After drying, the electrode was pressed with a line pressure of 1000 kgf. The loading level and the electrode thickness were determined to be about 2.27 ~ 2.79 mg cm-2 and 22 ~ 24 μm, respectively.

*Preparation and electrochemical test of cells*: Coin cells (2032) for charge-discharge tests were fabricated by sequentially superimposing the composite electrode, a porous polyethylene separator (Celgard), and finally the lithium metal (Honjo). The liquid electrolyte solution of 1.15 M $LiPF_6$ dissolved in 3:7 v/v mixture of ethylene carbonate and ethyl methyl carbonate was supplied by Enchem Co. Ltd.

Charge and discharge tests of the coin cells were carried out in the constant-current mode using a galvanostatic cycler (Toscat 3000, Toyo Systems) in a voltage range of 0.01-1 V with 0.1 C-rate (equivalent to 37.2 mA $g^{-1}_{NG}$). Cell impedance was measured in the frequency range of $10^{-1}$ to $10^6$ Hz with an amplitude of 10 mV using a frequency response analyzer (VSP, Biologic). The coin cells were operated at 27 °C.

*Surface characterizations*: After electrochemical reactions of the coin cells, the cells were disassembled and the composite electrodes were rinsed with dimethyl carbonate solvent followed by drying under vacuum for 12 h to remove the residual electrolyte species. In order to analyze the spatial distribution of components in electrode samples, we used a commercial atomic force microscope (AFM, MFP-3D Origin, Asylum research). All AFM analyses in this study were conducted at 25 °C under ambient air conditions. A conductive cantilever with a Cr/Pt overall-coated silicon tip (ContE-G, Budget sensors) was used for each measurement.

The cantilever has a nominal spring constant of 0.2 N $m^{-1}$ and a tip resonance frequency of 13 kHz for observing very small amplitude in amorphous SEI layers. The typical scan rate used for the measurements was 1 Hz and the scan angle was 0 °. The AFM topography images were obtained simultaneously during each AFM imaging mode. ESM imaging was performed to measure the electromechanical response of each sample.

In order to magnify the sensitivity of the measurement, we used dual-ac resonance tracking (DART) mode for ESM mapping. The AC drive voltage (2 V) with contact resonance frequency at around from 70 to 90 kHz was applied via the AFM tip while the tip loading force was 250 nN. We conducted each experiment at least 10 times to ensure repeatability.

The scanning electron microscopy (SEM, EC SNE-4500M) and energy-dispersive X-ray spectroscopy (EDS, Bruker XFlash 640H Mini) were used for surface structural analysis and elemental mapping, respectively. The X-ray photoelectron spectroscopy (XPS) tests (Al K-α, Thermo VG Scientific) were performed for surface chemical analysis of cycled electrodes. After scanning of survey and surface measurement, depth profile was carried out from 2 keV of $Ar^+$ ion beam for every 30 s. All spectra were calibrated by setting the C 1s photoemission peak for $sp^2$-hybridized carbon to 284.5 eV.

**Supporting Information**

Supporting Information is available from the Wiley Online Library or from the author.

**Acknowledgements**

This work was supported by the Materials Innovation Initiative Project of National Research Foundation (NRF) funded by the Korean Ministry of Science & ICT (2020M3H4A3081880) and the KAIST-funded Global Singularity Research Program for 2021 and 2022, and the Advanced Technology Research Center (ATC) Program (20001370) funded by the Ministry of Trade, Industry & Energy (MOTIE, Korea).

Jimin Oh, Gun Park, Hongjun Kim, Sujung Kim, Dong Ok Shin, Kwang Man Kim, Hye Ryung Byon, Young-Gi Lee, and Seungbum Hong*

**Correlating Nanoscale Structure with Electrochemical Property of Solid Electrolyte Interphases in Solid-State Battery Electrodes**

ToC figure ((Please choose one size: 55 mm broad × 50 mm high **or** 110 mm broad × 20 mm high. Please do not use any other dimensions))

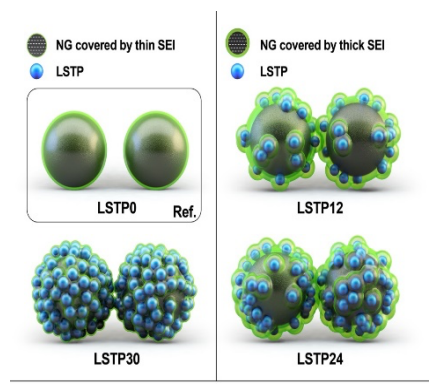

Supporting Information

**Correlating Nanoscale Structure with Electrochemical Property of Solid Electrolyte Interphases in Solid-State Battery Electrodes**

*Jimin Oh, Gun Park, Hongjun Kim, Sujung Kim, Dong Ok Shin, Kwang Man Kim, Hye Ryung Byon, Young-Gi Lee, and Seungbum Hong\**

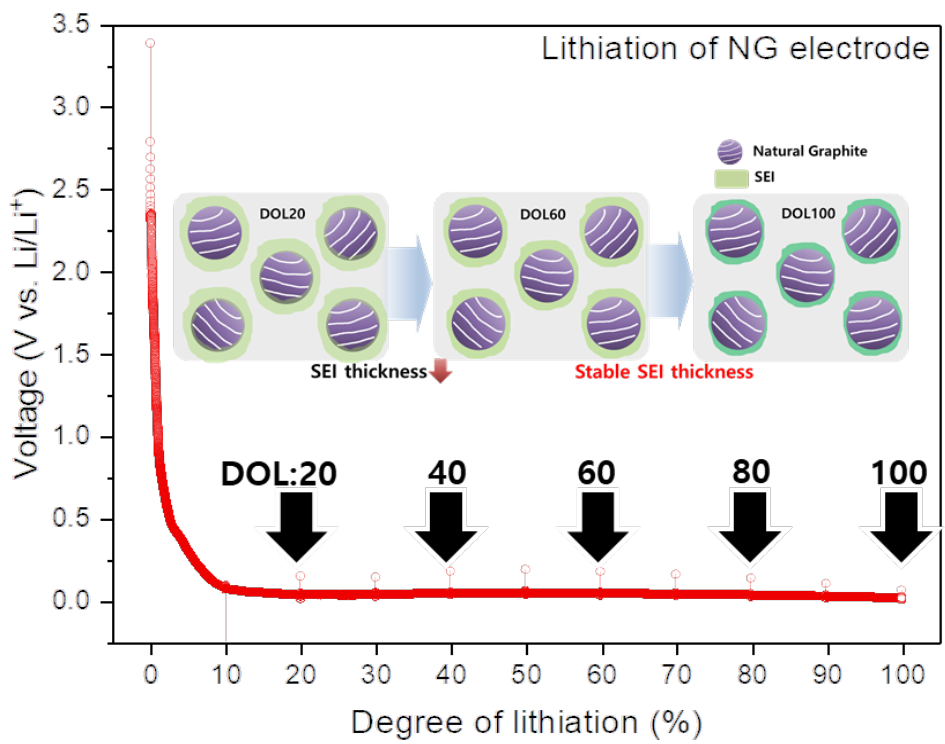

**Figure S1**. Lithiation profiles of NG electrode with different DOLs.

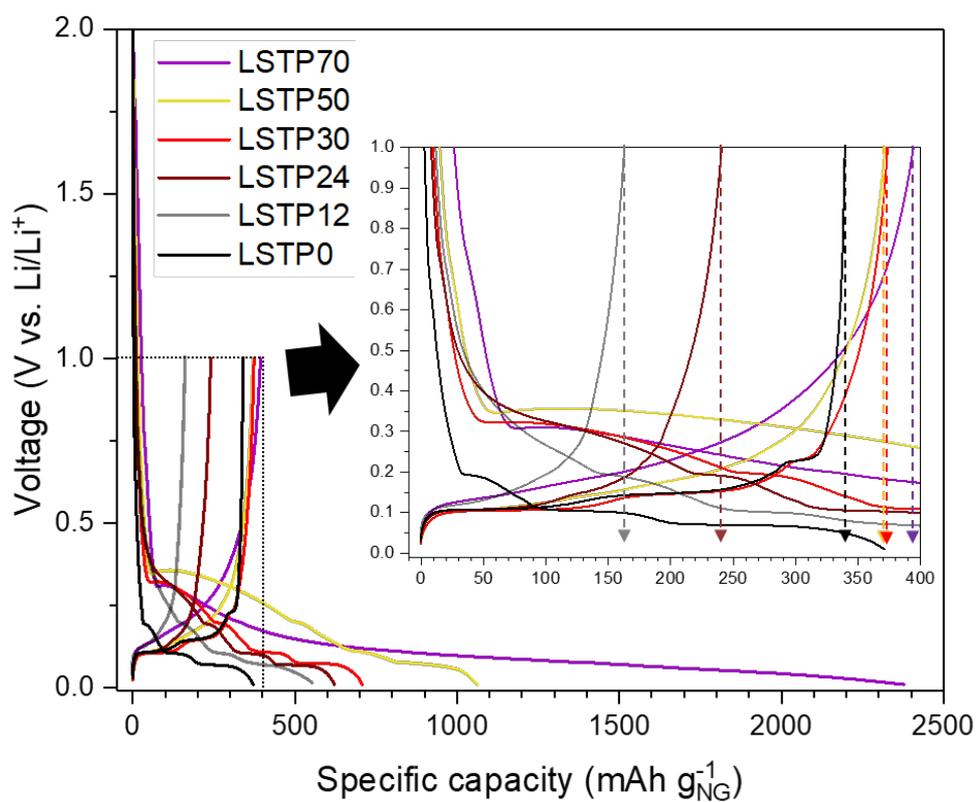

**Figure S2**. Voltage and specific capacity, as function of LSTP contents.

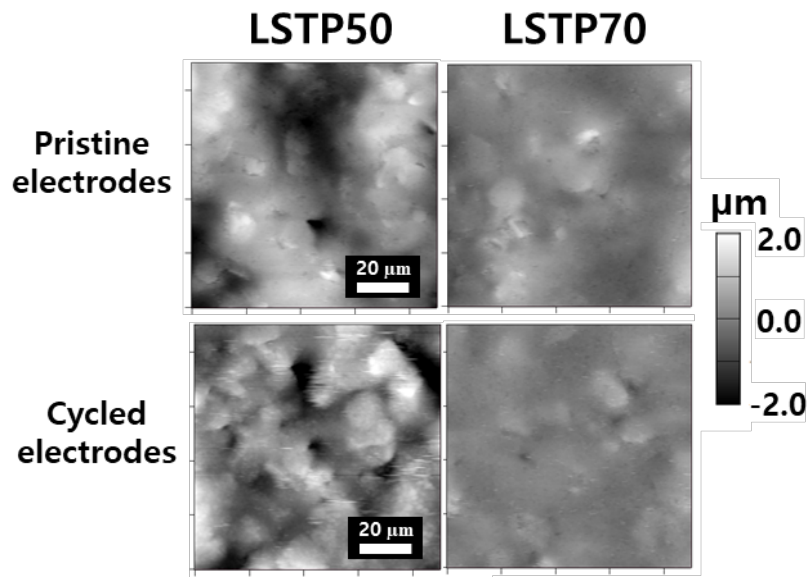

**Figure S3**. Height maps of pristine and 1 cycled electrode in LSTP50 and LSTP70.

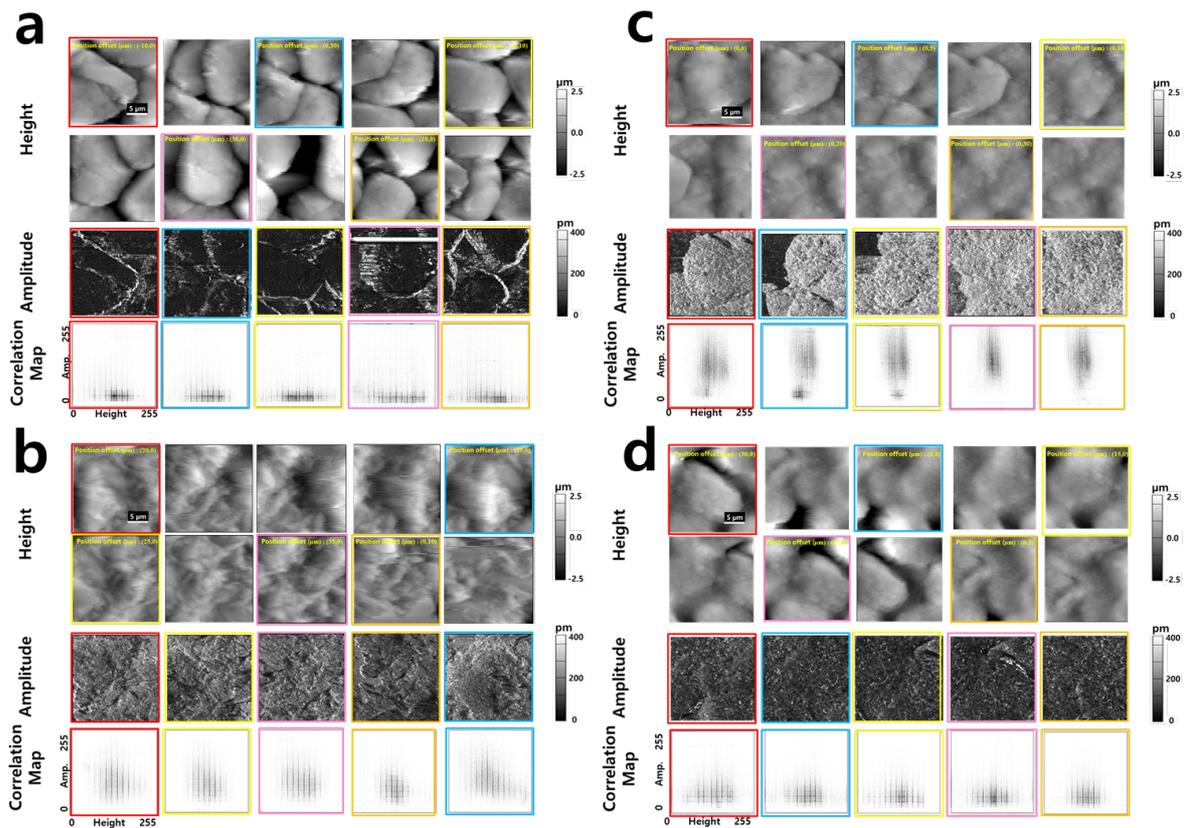

**Figure S4**. Height, amplitude at different 10 positions, and correlation maps at selected 5 positions from height, amplitude images (colored images) of 1 cycled electrodes in a) Ref. (LSTP0), b) LSTP12, c) LSTP24, and d) LSTP30.

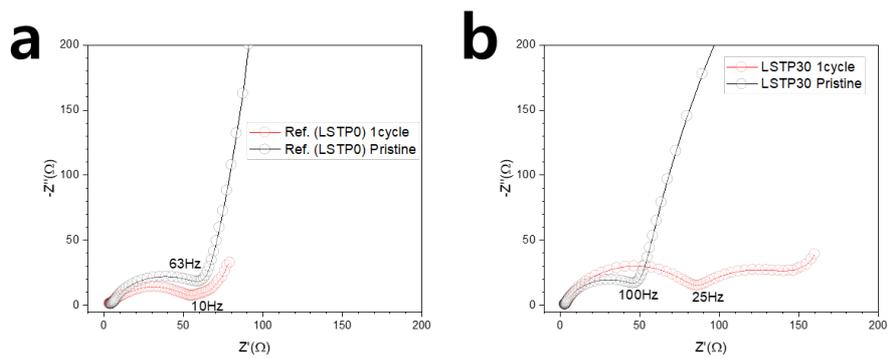

**Figure S5.** EIS results of pristine and 1cycle electrodes in a) Ref. (LSTP0) and b) LSTP30.

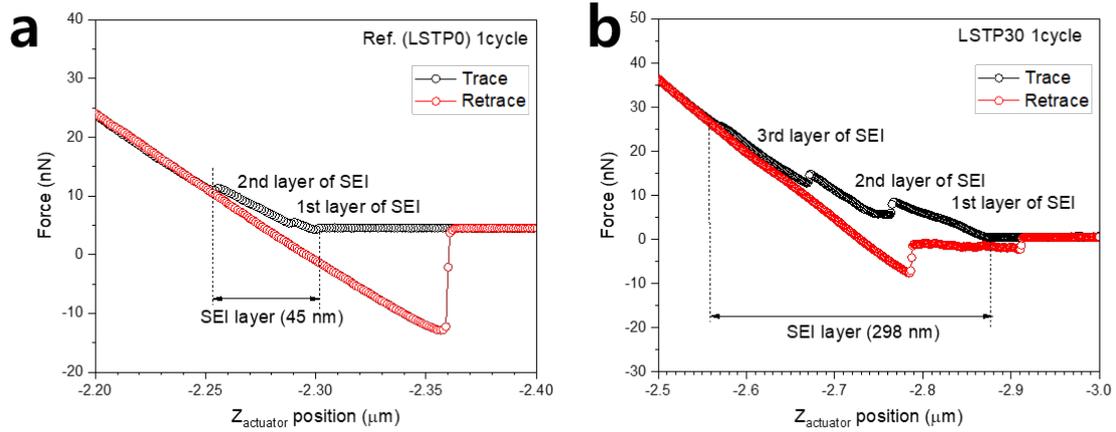

**Figure S6**. Force-Z actuator position curves of 1 cycled electrode in a) Ref. (LSTP0) and b) LSTP30 (spring constant $k$ = 0.164 N m$^{-1}$). In reference electrode, 2-layer SEI is observed with 45 nm, while in LSTP30 electrode, 3-layer SEI is observed with 298 nm.

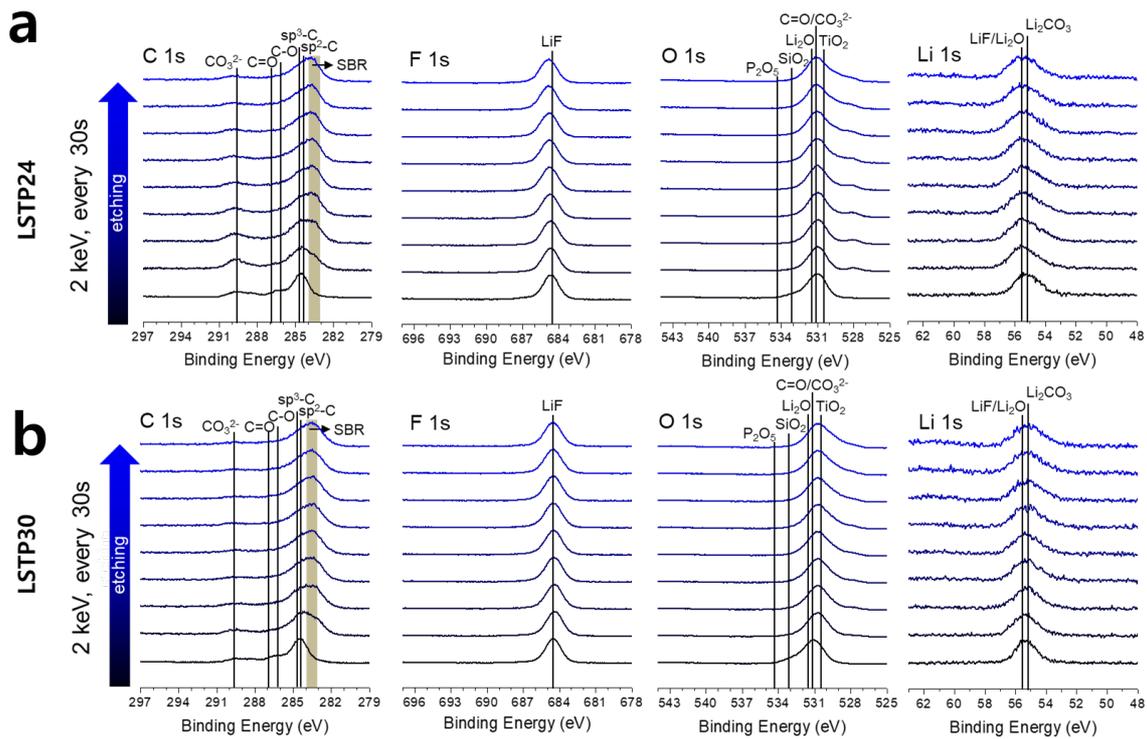

**Figure S7**. Photoelectron signals during argon etching in C 1s, F 1s, O 1s and Li 1s signals in 1 cycled a) LSTP24 and b) LSTP30 electrodes.

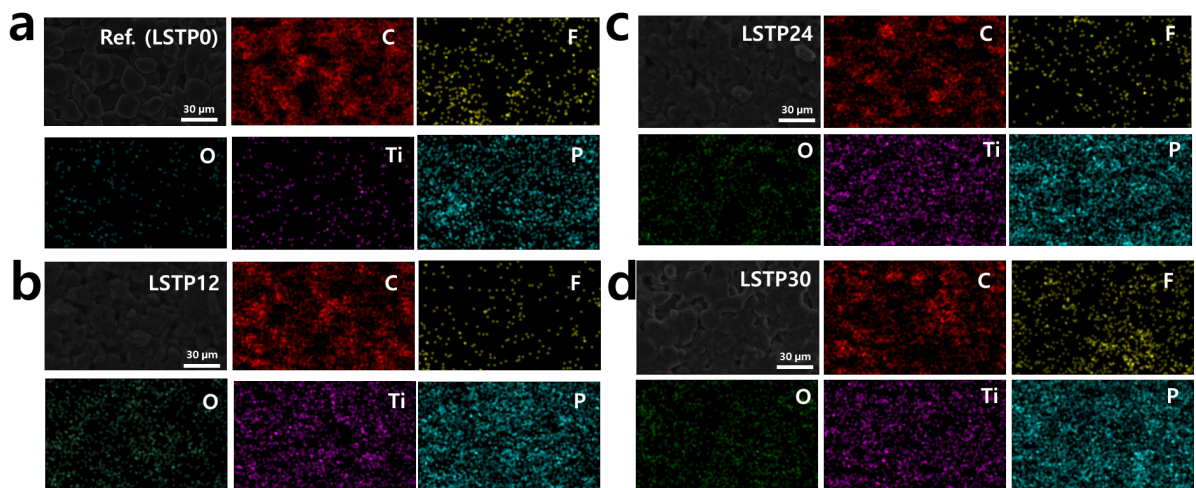

**Figure S8.** SEM/EDS results of 1 cycled electrodes in a) Ref. (LSTP0), b) LSTP12, c) LSTP24, and d) LSTP30 on the surface.

**Table S1**. Depth distributions in the lithium and fluorine

| [Li] | max([Li])-min([Li]) | [a)][Li]$_{surface}$ | max([Li])-min([Li]) / [a)][Li]$_{surface}$ |
|---|---|---|---|
| LSTP24 | 4.11 | 38.73 | 0.106 |
| LSTP30 | 2.25 | 38.28 | 0.059 |

| [F] | max([F])-min([F]) | [a)][F]$_{surface}$ | max([F])-min([F]) / [a)][F]$_{surface}$ |
|---|---|---|---|
| LSTP24 | 8.79 | 14.28 | 0.615 |
| LSTP30 | 10.03 | 20.14 | 0.498 |

[a)] [X]$_{surface}$ means the atomic percentage of X element at etching time 0 s